\documentclass[preprintnumbers,floats,twocolumn,prd,aps]{revtex4}
\usepackage{latexsym}
\usepackage{amsfonts}
\usepackage{psfrag}
\usepackage{graphicx}
\usepackage{bm}
\usepackage{amssymb,amsmath,epsfig}
\usepackage{subfig}
\usepackage{float}

\begin{document}

\setlength{\pdfpagewidth}{8.5in}

\setlength{\pdfpageheight}{11in}

\title{Bootstrapping Time Dilation Decoherence} 
\author{Cisco Gooding}\email{cgooding@physics.ubc.ca}
\author{William G. Unruh}\email{unruh@physics.ubc.ca}
\affiliation{
  Department of Physics and Astronomy,
  University of British Columbia\\
  Vancouver, British Columbia, V6T 1Z1 Canada
}
\date{\today}

\begin{abstract}
We present a general relativistic model of a spherical shell of matter with a perfect fluid on its surface coupled to an internal oscillator, which generalizes a model recently introduced by the authors to construct a self-gravitating interferometer \cite{GoodingUnruh14}. The internal oscillator evolution is defined with respect to the local proper time of the shell, allowing the oscillator to serve as a local clock that ticks differently depending on the shell's position and momentum. A Hamiltonian reduction is performed on the system, and an approximate quantum description is given to the reduced phase space. If we focus only on the external dynamics, we must trace out the clock degree of freedom, and this results in a form of intrinsic decoherence that shares some features with a proposed ``universal'' decoherence mechanism attributed to gravitational time dilation \cite{PZCB13}. We note that the proposed decoherence remains present in the (gravity-free) limit of flat spacetime, emphasizing that the effect can be attributed entirely to proper time differences, and thus is not necessarily related to gravity. Whereas the effect described in \cite{PZCB13} vanishes in the absence of an external gravitational field, our approach bootstraps the gravitational contribution to the time dilation decoherence by including self-interaction, yielding a fundamentally gravitational intrinsic decoherence effect.
\end{abstract}

\maketitle

\section{Introduction}

Time dilation is one of the most profound consequences of relativity theory. In classical systems, time dilation effects are fairly well understood, but in quantum systems there are still many questions that remain unanswered. One such question is how to properly incorporate the effects of time dilation into the quantum evolution of composite systems, either in the limit of flat spacetime or in situations where gravitational effects are significant. 

In a series of papers by Pikovski et al. \cite{PZCB13}-\cite{ZCPRB12}, for instance, a ``universal'' decoherence mechanism was proposed for composite general relativistic systems, due to gravitational time dilation. In \cite{PZCB13}, Pikovski et al. present an approximate quantum description of such a composite system that as a whole behaves as a point particle (located at the center-of-mass of the system) with a well-defined proper time. The system is placed in the gravitational field of the earth, and has internal degrees of freedom that are defined in the rest frame of the system. Expressing the proper time derivative in terms of the lab-frame time $t$ induces a coupling between the internal degrees of freedom and the center-of-mass coordinate, and if one only keeps track of the center-of-mass dynamics, tracing out over the internal degrees of freedom leads to a novel form of the ``third-party decoherence'' described by Stamp \cite{Stamp06}. It is this effect, and variations on the theme, that we focus on in this paper.

Rather than make use of the same model used by Pikovski et al., we will explore the same ideas with a model that generalizes the self-gravitating spherical perfect fluid shell introduced recently by the authors \cite{GoodingUnruh14}. Whereas our original model was introduced to study the consequences of general relativity on massive interferometers, here we extend the model to include an ``internal'' harmonic oscillator, to analyze the quantum structure of composite relativistic systems. By ``internal,'' we mean that the harmonic oscillator is described by an internal coordinate $q$ that oscillates in an abstract space that is not part of the spacetime; such an internal degree of freedom could represent the values of a single spherically symmetric mode of an oscillating field confined to the surface of the shell, for instance. The internal coordinate oscillates harmonically with respect to the proper time of the (external) shell position, and therefore the oscillator serves as a clock, evolving based on the local flow of time determined by the external motion.

 We should keep in mind that the internal oscillator contributes to the external shell dynamics as well, which in turn affects the spacetime; in other words, the very ticking of our clock influences the manner in which it ticks. This is especially relevant in the quantized system, because uncertainties in clock readings become intimately connected with uncertainties in spacetime geometry. 

We will explore some of the ambiguities associated with the quantum theory of this generalized shell system in reduced phase space, and then relate an approximate form of our reduced Hamiltonian with the Hamiltonian presented in \cite{PZCB13}. We exploit this parallel to demonstrate time dilation decoherence in our system. Even in the (gravity-free) limit of flat spacetime, we observe that when the fluid pressure is nonzero, decoherence (in the position basis) remains present, because of the acceleration caused by the pressure. This is a reflection of the fact that the effect proposed by Pikovski et al. results from proper time differences alone, and as such is not necessarily related to gravity. 

In the general case, our shell model includes gravitational self-interaction corrections to the time dilation decoherence, such that the decoherence is altered by the manner in which our shell and its clock influence the state of their own geometry. We find that even without pressure, the self-gravitation of the shell leads to the nonzero acceleration required to produce the position-basis time dilation decoherence. We interpret this ``self-decoherence'' as a fundamentally gravitational effect.

\section{Classical Action}

For context, before adding an internal oscillator, the self-gravitating spherical perfect fluid shell model introduced in \cite{GoodingUnruh14} is described by the action $I_x+I_G$, where $I_x$ is the shell action
\begin{equation}
I_{x}=- \int{d\lambda \,\sqrt{-g_{\mu\nu}\frac{d x^\mu}{d\lambda}\frac{d x^\nu}{d\lambda}}}M(R),
\label{eq:ShellAction}
\end{equation}
with all quantities evaluated on the shell history, and $I_G$ is the Einstein-Hilbert action
\begin{equation}
I = \frac{1}{16\pi}\int{d^4x\,\sqrt{-g^{(4)}} \hspace{3pt} \mathcal{R}^{(4)}}.
\end{equation} 
Here superscripts on the metric determinant $g$ and the Ricci scalar $\mathcal{R}$ indicate that these are constructed from the full $(3+1)$-dimensional spacetime metric components $\{g_{\mu\nu}\}$. We make use of the ADM form of the metric in spherical symmetry,
\begin{equation}
g_{\mu\nu}dx^{\mu}dx^{\nu}=-N^2 dt^2 + L^2\left(dr+N^r dt\right)^2+R^2d\Omega^2,
\label{eq:Metric}
\end{equation}
where $N$ is the lapse function, $N^r$ is the radial component of the shift vector, and $L^2$ and $R^2$ are the only nontrivial components of the spatial metric \cite{ADM}. It is then clear that $R$ is the ``radius'' of the shell, obtained from the area $4\pi R^2$ of symmetry two-spheres. The shell contribution $I_x$ is analogous to a free relativistic particle action, except with a mass $M$ that depends on the position-dependent metric function $R$; the function $M(R)$ serves to parametrize the relationship between the density $\sigma=M(R)/4\pi R^2$ and pressure $P_\sigma=-M'(R)/8\pi R$ of the fluid. 

We add an internal oscillator to our shell with the action
\begin{equation}
I_q=\frac{1}{2}\int{d\tau\, \left[m\left(\frac{dq}{d\tau}\right)^2-k q^2\right]},
\label{eq:Action}
\end{equation}
with $\tau$ being the proper time evaluated on the shell history, and $q$ being an internal coordinate that does not take values in the (external) spacetime. The quantity $k$ is related to $\omega_0$, the natural frequency of the oscillator, via $k=m\omega_0^2$. The action (\ref{eq:Action}) is manifestly invariant under coordinate transformations, as it only makes use of the proper time of the shell.

It simplifies the description to parametrize the shell history with the coordinate time $t$, such that the classical shell motion is defined by a trajectory $r=X(t)$. We can use the shell $4$-velocity $u^\mu=(d t / d \tau)(1,\dot{X},0,0)$ to express the proper time differentials as
\begin{equation}
d\tau=-u_\mu dx^\mu = \dot{\tau}dt
\end{equation}
and
\begin{equation}
\frac{dq}{d\tau}=-u^\mu \partial_\mu q= \dot{\tau}^{-1}\frac{dq}{dt}.
\end{equation}
Now $q$ is being treated as a function solely of coordinate time $t$, to reflect our choice of parametrization. The $4$-velocity normalization $u^\mu u_\mu=-1$ then implies that one can express the derivative of the proper time with respect to the coordinate time as
\begin{eqnarray}
\dot{\tau}&=&\int{dr\,\sqrt{N^2-L^2(N^r+\dot{X})^2}\delta(r-X)}\nonumber\\
&=&\sqrt{\hat{N}^2-\hat{L}^2(\hat{N}^r+\dot{X})^2}.
\end{eqnarray}
An overhat denotes that a quantity is to be evaluated on the shell history; likewise, it is understood that the overdots denote coordinate-time derivatives along the shell trajectory.

If we define the original shell Lagrangian as
\begin{eqnarray}\label{ShellLq}
\mathcal{L}_x&=&-\int{dr\,\sqrt{N^2-L^2(N^r+\dot{X})^2}M(R)\delta(r-X)} \nonumber\\
&=& -\dot{\tau}\hat{M}
\end{eqnarray}
and the oscillator Lagrangian as
\begin{eqnarray}\label{ClockLq}
\mathcal{L}_q&=&\frac{1}{2}\int{dr\, \dot{\tau}\left[m\left(\frac{\dot{q}}{\dot{\tau}}\right)^2-k q^2\right]\delta(r-X)}\nonumber\\
&=& \frac{1}{2}\left(m\frac{\dot{q}^2}{\dot{\tau}}-k\dot{\tau}q^2\right),
\end{eqnarray}
then the shell-oscillator action is given by
\begin{equation}
I_{shell}=\int dt\,\mathcal{L}=\int{dt\,\left(\mathcal{L}_x+\mathcal{L}_q\right)}.
\end{equation}

\section{Hamiltonianization}

We can Hamiltonianize the shell-oscillator system with the Legendre transformation $\mathcal{H}=P\dot{X}+p\dot{q}-\mathcal{L}$. The momentum $p$ conjugate to the internal coordinate $q$ is given by
\begin{equation}
p \equiv \frac{\partial\mathcal{L}}{\partial\dot{q}}=m\int{dr\,\frac{\dot{q}}{\dot{\tau}}\delta(r-X)}=m\frac{\dot{q}}{\dot{\tau}},
\end{equation}
and that the momentum conjugate to the shell position $X$ is given by
\begin{eqnarray}
P&\equiv&\frac{\partial \mathcal{L}}{\partial \dot{X}}=-\hat{M}\frac{\partial\dot{\tau}}{\partial\dot{X}}-\frac{1}{2}m\left(\frac{\dot{q}}{\dot{\tau}}\right)^2\frac{\partial\dot{\tau}}{\partial\dot{X}}-\frac{1}{2}kq^2\frac{\partial\dot{\tau}}{\partial\dot{X}}\nonumber\\
&=&-\frac{\partial\dot{\tau}}{\partial\dot{X}}\left(\hat{M}+\frac{p^2}{2m}+\frac{1}{2}kq^2\right)\nonumber\\
&=&-\frac{\partial\dot{\tau}}{\partial\dot{X}}\left(\hat{M}+H_q\right),
\end{eqnarray}
where  the internal ``clock'' Hamiltonian $H_q=p^2/2m+kq^2/2$ takes the form of a (free) harmonic oscillator. Introducing the notation $\tilde{M}=\hat{M}+H_q$, one can observe that our shell-oscillator system becomes very similar to the shell system without the oscillator, subject to the transformation $\hat{M}\rightarrow\tilde{M}$. 

More explicitly, the shell momentum $P$ can be expressed as
\begin{equation}
P=\int{dr\,\frac{L^2\left(N^r+\dot{X}\right)\tilde{M}}{\sqrt{N^2-L^2\left(N^r+\dot{X}\right)^2}}\delta\left(r-X\right)},
\end{equation}
from which we can solve for $\dot{X}$ to obtain
\begin{eqnarray}
\dot{X}&=&\int{dr\,\left(\frac{NP}{L\sqrt{P^2+L^2\tilde{M}^2}}-N^r\right)\delta\left(r-X\right)}\nonumber\\
&=& \frac{\hat{N}P}{\hat{L}\sqrt{P^2+\hat{L}^2\tilde{M}^2}}-\hat{N}^r.
\end{eqnarray}
The Legendre transformation then gives us the Hamiltonian
\begin{equation}
\mathcal{H}=P\dot{X}+p\dot{q}-\mathcal{L}=\int{dr\, \left(NH_t^s+ N^rH_r^s\right)},
\end{equation}
with the definitions
\begin{eqnarray}
H_t^s &=& \sqrt{L^{-2}P^2+\tilde{M}^2}\delta(r-X), \nonumber \\ 
H_r^s &=& -P\delta(r-X).
\end{eqnarray}
We remind the reader that $\tilde{M}=\hat{M}+H_q$, so our clock Hamiltonian adds to the (position-dependent) shell mass $\hat{M}$ to alter the Hamiltonian constraint from the form it took in \cite{GoodingUnruh14}.

Hamiltonianizing the gravitational sector as well leads to the total action
\begin{eqnarray}
I =&& \int{dt\, \left(P\dot{X}+p\dot{q}\right)}\nonumber\\
&&+\int{dt\,dr\,\left(\pi_R \dot{R}+\pi_L \dot{L}-NH_t-N^r H_r\right)},
\end{eqnarray}
for $H_t=H_t^s+H_t^G$ and $H_r=H_r^s+H_r^G$, such that
\begin{eqnarray}
H_t^G &=& \frac{L\pi_L^2}{2R^2}-\frac{\pi_L\pi_R}{R}+\left(\frac{R R'}{L}\right)'-\frac{(R')^2}{2L}-\frac{L}{2}, \nonumber\\
H_r^G &=& R'\pi_R-L\pi_L'.
\label{eq:GravConstraints}
\end{eqnarray}

\section{Reduced Phase Space Quantization}

Since we are working in spherical symmetry, the metric itself has no actual degrees of freedom, because although there are only two gravitational constraints ($H_t=0$ and $H_r=0$), there are also only two independent metric functions ($L$ and $R$). Accordingly, one can obtain an unconstrained description of the system by making a coordinate choice, solving the constraints for the corresponding gravitational momenta, and inserting the solutions into the Liouville form $\mathcal{F}$ on the full phase space,
\begin{eqnarray}
\mathcal{F}=P\bm{\delta} X +p\bm{\delta} q+ \int{dr\left(\pi_L \bm{\delta} L +\pi_R \bm{\delta} R\right)}.
\end{eqnarray}
This amounts to a pullback of the full Liouville form to the representative hypersurface defined by the coordinate choice. From the Liouville pullback, denoted by $\tilde{\mathcal{F}}$, we can deduce the canonical structure of the reduced phase space, which only depends on the shell-oscillator variables $X$ and $q$ (and their momenta).

To solve the gravitational constraints, first consider the following linear combination of the constraints, away from the shell:
\begin{equation}
-\frac{R'}{L}H_t-\frac{\pi_L}{RL}H_r=\mathcal{M}',
\end{equation}
for
\begin{equation}
\mathcal{M}(r)=\frac{\pi_L^2}{2R}+\frac{R}{2}-\frac{R(R')^2}{2L^2}.
\end{equation}
The quantity $\mathcal{M}(r)$ corresponds to the ADM mass $H$ when evaluated outside of the shell, and vanishes inside the shell. We can now solve for the gravitational momenta $\pi_L$, $\pi_R$ away from the shell. The result is
\begin{equation}
\pi_L=\pm R\sqrt{\left(\frac{R'}{L}\right)^2-1+\frac{2\mathcal{M}}{R}},\hspace{8pt}\pi_R=\frac{L}{R'}\pi_L'.
\label{eq:GravMomenta}
\end{equation}

Assuming a continuous metric and singularity-free gravitational momenta, we can integrate the gravitational constraints ($H_t=0$ and $H_r=0$) across the shell, from which we obtain the jump conditions
\begin{equation}
\Delta R'=-\frac{\tilde{V}}{\hat{R}}, \hspace{8pt} \Delta\pi_L=-\frac{P}{\hat{L}},
\label{eq:Jumpsq}
\end{equation}
where $\tilde{V}=\sqrt{P^2+\tilde{M}^2}$. We use $\Delta$ to denote the jump of a quantity across the shell (at $r=X(t)$). 

The coordinates we will use resemble the Painlev\'e-Gullstrand coordinates $\{L=1,R=r\}$, though the jump conditions force us to include a deformation region ($X-\epsilon<r<X$) near the shell. By inspection, the required metric function $R$ can be generalized as
\begin{equation}
R(r,t)=r-\frac{\epsilon}{X}\tilde{V}\mathcal{G}\left(\frac{X-r}{\epsilon}\right),
\label{eq:Coordsq}
\end{equation}
for a function $\mathcal{G}$ having the properties
\begin{align}
 \lim_{z\rightarrow 0^+}\frac{d\mathcal{G}(z)}{dz}&=1\\
 \lim_{z\rightarrow 0^-}\frac{d\mathcal{G}(z)}{dz}&=0\,,
\end{align}
from which it follows that
\begin{align}
&\lim_{\epsilon\rightarrow 0}R' (X-\epsilon)=1+\frac{\tilde{V}}{X}\\
&\lim_{\epsilon\rightarrow 0}R' (X+\epsilon)=1\,.
\end{align}

By inserting the gravitational momentum solutions (\ref{eq:GravMomenta}) associated with the coodinate choice (\ref{eq:Coordsq}) into the jump equations (\ref{eq:Jumpsq}) and squaring, one finds
\begin{equation}
H=\sqrt{P^2+\tilde{M}^2}+\frac{\tilde{M}^2}{2X}-P\sqrt{\frac{2H}{X}}.
\label{eq:DeterminesPq}
\end{equation} 
This implies that the unreduced momentum $P$ is implicitly defined as a function of the reduced phase space quantities $X$, $q$, $p$, and $H$: one can easily solve (\ref{eq:DeterminesPq}) to obtain
\begin{eqnarray}\label{eq:Pexplicitq}
P &=& \frac{1}{1-\frac{2H}{X}}\left(\sqrt{\frac{2H}{X}}\left(H-\frac{\tilde{M}^2}{2X}\right)\right) \\
&&\pm \frac{1}{1-\frac{2H}{X}}\left(\sqrt{\left(H-\frac{\tilde{M}^2}{2X}\right)^2-\tilde{M}^2\left(1-\frac{2H}{X}\right)}\right) \nonumber .
\end{eqnarray}
The $\pm$ in (\ref{eq:Pexplicitq}) indicates whether the shell is outgoing ($+$) or ingoing (-), with respect to our choice of coordinates.

Let us now calculate the pullback of the full Liouville form to the representative hypersurface defined by our coordinate choice. The condition $L=1$ and the fact that $R=r$ outside of the deformation region implies that
\begin{equation}\label{eq:LiouvillePullq}
\tilde{\mathcal{F}}=P\bm{\delta} X+p\bm{\delta} q+
\int_{X-\epsilon}^X dr\, \pi_R \bm{\delta} R.
\end{equation}
We can then simplify the remaining integral by changing the integration variable from $r$ to $v=R'$, which yields
\begin{equation}
\int_{X-\epsilon}^{X}{dr\, \pi_R \bm{\delta} R}=X\bm{\delta} X\int_{1}^{R_-'}{dv\,\frac{\left(1-v\right)}{\sqrt{v^2-1}}}+\mathcal{O}(\epsilon),
\end{equation}
with $R_-'$ being $R'$ evaluated just inside the shell. Now the integration is trivial, and we can easily obtain the desired Liouville form pullback,
\begin{equation}
\mathcal{F}=P_c \bm{\delta} X+p \bm{\delta}q,
\end{equation}
with the reduced canonical momentum for the shell position satisfying
\begin{eqnarray}
P_c = -\sqrt{2HX}+X\ln{\left(1+\frac{\tilde{V}+P}{X}+\sqrt{\frac{2H}{X}}\right)}.
\label{eq:Pcq2}
\end{eqnarray}
This expression gives an implicit definition of the Hamiltonian $H$ on the reduced phase space, as a function of the shell-oscillator variables ($X$ and $q$), along with the momenta that are conjugate to them in the reduced phase space ($P_c$ and $p$, respectively).

It can also be shown \cite{GoodingUnruh14} that the expression (\ref{eq:Pcq2}) for the reduced canonical shell momentum $P_c$ is equivalent to the form analogous to the expression given by Kraus and Wilczek \cite{KrausWilczek},
\begin{equation}
P_c = -\sqrt{2HX}-X\ln{\left(\frac{X+\tilde{V}-P-\sqrt{2HX}}{X}\right)},
\label{eq:Pcq}
\end{equation}
despite the different minus sign placement.

To gain some intuition for how the presence of the oscillator alters the reduced dynamics, let us consider the flat spacetime limit ($X\rightarrow\infty$) of the system defined by (\ref{eq:Pcq}). Keeping in mind the similarities with the relativistic-particle-like structure of our shell system, it should be unsurprising that in this limit (\ref{eq:Pcq}) becomes
\begin{equation}
P_c=\pm\sqrt{H^2-\tilde{M}^2}=\pm\sqrt{H^2-\left(\hat{M}+H_q\right)^2},
\end{equation}
and therefore the Hamiltonian is given by
\begin{equation}
H=\sqrt{P_c^2+\tilde{M}^2}=\sqrt{P_c^2+\left(\hat{M}+H_q\right)^2}.
\end{equation}
In the nonrelativistic regime (i.e. small $P_c$), the Hamiltonian can then be expressed as
\begin{eqnarray}\label{HQFlat}
H&\approx&\hat{M}+H_q+\frac{P_c^2}{2\left(\hat{M}+H_q\right)}\nonumber\\
&=&\hat{M}+\frac{1}{2}kq^2+\frac{p^2}{2m}+\frac{P_c^2}{2\left(\hat{M}+\frac{1}{2}kq^2+\frac{p^2}{2m}\right)}.
\end{eqnarray}
The last term in this approximate Hamiltonian is an effective coupling between the internal oscillator variables ($q$ and $p$) and the external shell variables ($X$ and $P_c$). The coupling is of course produced by the fact that the internal ``clock'' oscillates harmonically with the shell's proper time, the flow of which is influenced by the external variables.

Even in the flat spacetime nonrelativistic limit, one can tell from the appearance of the clock Hamiltonian $H_q$ in the denominator of the last term in (\ref{HQFlat}) that exact quantization will require nonstandard techniques. The Hamiltonian (\ref{HQFlat}) leads to the following Schr\"odinger equation, in the coordinate basis:
\begin{eqnarray}\label{SchrodQ}
i\frac{\partial}{\partial t}\Psi &=&\left(\hat{M}+\frac{1}{2}kq^2-\frac{1}{2m}\frac{\partial^2}{\partial q^2}\right)\Psi\nonumber\\
&&-\frac{1}{2}\frac{\partial}{\partial X}\left[\frac{1}{\hat{M}+\frac{1}{2}kq^2-\frac{1}{2m}\frac{\partial^2}{\partial q^2}}\right]\frac{\partial}{\partial X}\Psi.
\end{eqnarray}
The factor-ordering in the last term of (\ref{SchrodQ}) was chosen to make the differential operator Hermitian, but there is still some ambiguity in the meaning of the bracketed factor between the $X$-derivatives, since the formal expression has $q$-derivatives in the denominator.

Instead of working directly with the Hamiltonian (\ref{HQFlat}), we will work with an approximate Hamiltonian that allows for a simpler analysis. Before doing so, however, let us consider the lowest-order corrections to (\ref{HQFlat}) due to gravitational self-interaction. This can be accomplished by inserting the ansatz $H=F_0(X)+F_1(X) P_c+F_2(X) P_c^2$ into (\ref{eq:Pcq}) and expanding for large $X$ and small $P_c$. If we assume that the $X$-dependence of $\tilde{M}$ does not significantly affect the large-$X$ behaviour of the functions $\{F_i(X)\}$, then we find \cite{GoodingUnruh14}
\begin{eqnarray}
F_0&=&\tilde{M}-\frac{\tilde{M}^2}{18 X}+\mathcal{O}(1/X^2)\nonumber\\
F_1&=&-\frac{2}{3}\sqrt{\frac{2\tilde{M}}{X}}+\mathcal{O}(1/X^{3/2})\\
F_2&=&\frac{1}{2\tilde{M}}+\frac{1}{3X}+\mathcal{O}(1/X^2)\,.\nonumber
\end{eqnarray}

We will now assume that the shell mass $\hat{M}$ is sufficiently larger than the clock energy $H_q$ that we can express $\tilde{M}$ as $\hat{M}\left(1+H_q/\hat{M}\right)$ and expand the Hamiltonian in powers of $H_q/\hat{M}$. To lowest order, the system then decomposes into a more standard form, given by
\begin{equation}\label{Decomp}
H=H_0+H_{xq}=H_x+H_q+H_{xq},
\end{equation}
with the approximate shell Hamiltonian 
\begin{eqnarray}\label{Hx}
H_x = \hat{M}+\frac{P_c^2}{2\hat{M}}+\left[\frac{P_c^2}{3X}-\frac{2}{3}\sqrt{\frac{2\hat{M}}{X}}P_c-\frac{\hat{M}^2}{18X}\right],
\end{eqnarray} 
the internal ``clock'' Hamiltonian $H_q=p^2/2m+kq^2/2$, and the interaction 
\begin{eqnarray}\label{Hxq}
H_{xq}=-\frac{P_c^2}{2\hat{M}^2}H_q+\left[-\frac{\hat{M}}{9X}H_q-\frac{1}{3\hat{M}}\sqrt{\frac{2\hat{M}}{X}}H_q\right],
\end{eqnarray}
which is induced by the clock oscillation being defined with respect to the proper time of the shell.

\section{Time Dilation Decoherence}

The decomposition (\ref{Decomp}) is of the same form as the one used recently by Pikovski et al. \cite{PZCB13} to demonstrate decoherence due to gravitational time dilation for composite systems, though the interpretation of the system variables is different. The main difference is that the ``external'' coordinate of our system is the shell radius instead of the (somewhat ill-defined) center-of-mass coordinate. In our case, however, the self-gravitation of the shell-plus-clock system is taken into account, so both the shell and the clock influence the spacetime geometry, which is therefore no longer fixed. Nonetheless, we can exploit the similarity enough to demonstrate an analogous decoherence in our system, as will become clear in what follows.

Represented by a density operator, the full state $\rho$ obeys the von Neumann equation,
\begin{equation}
\dot{\rho}=-i\left[H,\rho\right].
\end{equation}
We can then change the frame to primed coordinates, as in \cite{PZCB13}, which are defined by $\rho'(t)=e^{it(H_0+h)}\rho(t)e^{-it(H_0+h)}$, for $h(X,P_c)=Tr_q\left[H_{xq}\rho_q(0)\right]$. We are assuming that the initial state of the system is of the product form $\rho(0)=\rho_x(0) \rho_q(0)$, i.e. initially uncorrelated. Denoting the average clock energy $Tr_q\left[H_{q}\rho_q(0)\right]$ by $\bar{E}_q$ and the shell part of the interaction by 
\begin{equation}\label{Gamma}
\Gamma=-\frac{P_c^2}{2\hat{M}^2}+\left[-\frac{\hat{M}}{9X}-\frac{1}{3\hat{M}}\sqrt{\frac{2\hat{M}}{X}}P_c\right],
\end{equation}
we obtain the expression $h=\Gamma(X,P_c)\bar{E}_q$. In equation (\ref{Gamma}), as well as equations (\ref{Hx}) and (\ref{Hxq}), the bracketed term originates from self-gravitation. The transformed von Neumann equation is
\begin{eqnarray}\label{vonNeumann}
\dot{\rho}'(t)&=&i\left[H_0'(t)+h'(t),\rho'(t)\right]-i\left[H_0'(t)+H_{xq}'(t),\rho'(t)\right]\nonumber\\
&=&-i\left[H_{xq}'(t)-h'(t),\rho'(t)\right],
\end{eqnarray}
with $h'(t)=h(X'(t),P_c'(t))$. If we integrate and iterate equation (\ref{vonNeumann}), we are led to the integro-differential equation
\begin{eqnarray}
\dot{\rho}'(t)&=&-i\left[H_{xq}'(t)-h'(t),\rho'(0)\right]\\
&&-\int_0^t ds\,\left[\tilde{h}(t),\left[\tilde{h}(s),\rho'(s)\right]\right]\nonumber,
\end{eqnarray}
using the definition $\tilde{h}(t)=H_{xq}'(t)-h'(t)$. At this point Pikovski et al. trace over the internal variables, which for us describe the clock, and make use of the Born part of the Born-Markov approximation, keeping only terms up to second order in the interaction Hamiltonian $H_{xq}$, and replacing the $\rho'(s)$ in the integral by $\rho_x'(s)\rho_q'(0)$. This application of the Born approximation assumes weak coupling, but in contrast to the full Born-Markov it does not ignore memory effects. For a detailed discussion of this approximation, see \cite{WallsMilburn94}. The reduced equation for the shell variables is then given by
\begin{eqnarray}
&&\dot{\rho}_x'(t)=Tr_q\left[\dot{\rho}'(t)\right]\nonumber\\
&&\approx -\int_0^t ds\,Tr_q\left\{ \left[\tilde{h}(t),\left[\tilde{h}(s),\rho'(s)\right]\right]\right\}\nonumber\\
&&=-\int_0^t ds\, Tr_q\left\{\left(H_q-\bar{E}_q\right)^2\left[\Gamma'(t),\left[\Gamma'(s),\rho'(s)\right]\right]\right\}\nonumber\\
&&=-\left(\Delta E_q\right)^2\int_0^t ds\, \left[\Gamma'(t),\left[\Gamma'(s),\rho_{x}'(s)\right]\right],
\end{eqnarray}
with the notation $\Gamma'(s)=\Gamma(X'(s),P_c'(s))$ and 
\begin{equation}
\left(\Delta E_q\right)^2=Tr_q\left[\left(H_q-\bar{E}_q\right)^2\rho_q\left(0\right)\right].
\end{equation}
One can then transform back to the unprimed frame, whereby the substitution $s\rightarrow t-s$ 
leads to the expression
\begin{widetext}
\begin{eqnarray}\label{ClockMaster}
\dot{\rho}_x(t)=-i\left[H_x+\Gamma\bar{E}_q,\rho_x(t)\right]-\left(\Delta E_q\right)^2\int_0^t ds\, \left[\Gamma,e^{-isH_x}\left[\Gamma,\rho_x(t-s)\right] e^{isH_x}\right].
\end{eqnarray}
\end{widetext}

In general, the reduced evolution equation (\ref{ClockMaster}) exhibits decoherence due to the nonunitary contribution of the last term on the right. Under some special circumstances this term vanishes, leaving the reduced system to evolve unitarily; for example, such a circumstance occurs for initial states that are eigenstates of the internal (clock) Hamiltonian, of course then the oscillator is not much of a clock, as it never changes with time (modulo a phase).

\section{Discussion}

In the last section, we demonstrated intrinsic decoherence due to time dilation. The decoherence basis is in general a combination of position and momentum, reducing to purely position in the limit of negligible shell momentum. In the system described by Pikovski et al. \cite{PZCB13}, which includes spacetime curvature caused by the external gravitational field of the earth, the part of the time dilation decoherence that involves the position basis should vanish in the absence of the earth's gravitational influence: without the earth, the center-of-mass coordinate they use defines the origin of an inertial frame, and in that frame the proper time associated with the center-of-mass coordinate is equal to the coordinate time \cite{Comment}. However, in our system, this decoherence is present even in the limit of flat spacetime (ignoring both external gravitational fields and self-gravitation), because of the nonzero acceleration of the shell due to the position dependence of the mass ($\hat{M}=M(X)$). We then see a confirmation that the time dilation decoherence proposed in \cite{PZCB13} is not necessarily related to gravity, but produced by proper time differences in composite systems with nonzero accelerations.

Although a simpler form of the shell model we employ was previously used by the authors to examine the possible origins of the type of gravitational decoherence explored by Penrose \cite{Penrose} and Di\'osi \cite{Diosi}, the self-gravitation described here results from classical general relativity alone, and is thus not related to speculative models that produce self-gravitation by altering quantum mechanics. It is still an open question whether or not Penrose-type gravitational decoherence can be demonstrated within canonical quantum gravity, without introducing any new physics (for further discussion, see \cite{GoodingUnruh14}).

Another observation we can make is that even in the constant $\hat{M}$ limit, where the fluid pressure vanishes, equation (\ref{ClockMaster}) indicates that the time dilation decoherence remains present, in this case because the self-gravitation produces a nonzero acceleration of the shell position. Since the decohering term in the reduced evolution equation (\ref{ClockMaster}) varies with the square of the clock's energy uncertainty, increasing the energy uncertainty of the clock enhances the decoherence. In our system the clock energy contributes to the ADM energy of the spacetime, so uncertainties in the clock energy contribute to uncertainties in the spacetime geometry, which in turn lead to the type of self-decoherence mentioned in the introduction. Conceptually, such an effect should occur for any composite general relativistic system that has internal motion that (classically) evolves according to the proper time associated with the system's external motion, since the alteration of the local flow of time caused by the system's influence on its own spacetime geometry induces an effective coupling between the internal and external degrees of freedom. It is \textit{this} effect that is fundamentally gravitational in nature, as it is present even in the absence of any other interactions.

We have therefore arrived at a type of intrinsic decoherence similar to the ``third-party'' decoherence described by Stamp \cite{Stamp06}, though in contrast to the use of the earth as the third party as proposed by Pikovski et al. \cite{PZCB13}, we have bootstrapped the idea by incorporating gravitational self-interaction, effectively producing third-party decoherence without the third party. 

\section{Acknowledgements}

The authors would like to thank the Natural Sciences and Engineering Research Council of Canada (NSERC) and the Templeton Foundation (Grant No. JTF $36838$) for financial support. Also, we are grateful to the Aspelmeyer and Brukner groups at the University of Vienna, as well as Friedemann Queisser, Dan Carney, Philip Stamp, and Bob Wald, for stimulating discussions.


\begin{thebibliography}{99}

\bibitem{GoodingUnruh14}
  C.~Gooding and W.~G.~Unruh,
	``Self-gravitating interferometry and intrinsic decoherence,''
	Phys.\ Rev.\  D {\bf 90}, 044071 (2014).

	\bibitem{PZCB13}
I.~Pikovski, M.~Zych, F.~Costa, and \u{C}.~Brukner, 
``Universal decoherence due to gravitational time dilation,''
accepted for publication in Nature Physics, doi:10.1038/nphys3366 (2015).
[arXiv:quant-ph/1311.1095]

\bibitem{ZCPB11}
M.~Zych, F.~Costa, I.~Pikovski, and \u{C}.~Brukner,
``Quantum interferometric visibility as a witness of general relativistic proper time,''
Nat.\ Commun.\ {\bf 2}, 505 (2011).
[arXiv:quant-ph/1105.4531v2]

\bibitem{ZCPRB12}
M.~Zych, F.~Costa, I.~Pikovski, T.C.~Ralph and \u{C}.~Brukner,
``General relativistic effects in quantum interference of photons,''
	Class.\ Quantum Grav.\ {\bf 29}, 224010 (2012).
[arXiv:quant-ph/1206.0965v2]	
	
\bibitem{Stamp06}
P.~C.~E.~Stamp,
``The decoherence puzzle,''
Stud.\ in Hist.\ and Phil.\ of Mod. Phys.\ {\bf 37}, 467 (2006).	
	
\bibitem{ADM}
	R.~L.~Arnowitt, S.~Deser, and C.~W.~Misner,
  ``Canonical Variables for General Relativity,''
  Phys.\ Rev.\ {\bf 117}, 1595 (1960).
  [arXiv:gr-qc/0405109]	

\bibitem{KrausWilczek}
  P.~Kraus and F.~Wilczek,
  ``Self-Interaction Correction to Black Hole Radiance,''
	Nucl.\ Phys.\ B {\bf 433}, 403 (1995).
  [arXiv:gr-qc/9408003v1]
	
\bibitem{WallsMilburn94}
 D.~Walls and G.~Milburn,
 ``Quantum Optics,''
Springer-Verlag, Berlin (1994).		

\bibitem{Comment}
Of course, quantum fluctuations of the center-of-mass motion would still produce decoherence in the momentum basis for the reduced system; similarly, for our shell system in the absence of both gravity and pressure, the effective interaction (\ref{Hxq}) would lead to decoherence in the momentum basis of the reduced system. In both of these cases the decoherence basis would not involve the position, so it would be difficult to imagine how to observe such a decoherence using standard techniques such as interferometry.

\bibitem{Penrose}
R.~Penrose,
``On Gravity's role in Quantum State Reduction,''
Gen.\ Rel.\ Grav.\ {\bf 28}, 581-600 (1996).

\bibitem{Diosi}
L.~Di\'osi,
``Models for universal reduction of macroscopic quantum fluctuations,''
Phys.\ Rev.\ A {\bf 40}, 1165-1174 (1989).

\end{thebibliography}
\end{document}